\documentclass{JHEP3}

\usepackage{amsmath,amsfonts,amsthm,times,pst-node}

{\count255=\time\divide\count255 by 60
\xdef\hourmin{\number\count255}
\multiply\count255 by-60\advance\count255 by\time
\xdef\hourmin{\hourmin:\ifnum\count255<10 0\fi\the\count255}}
\def\ps@draft{\let\@mkboth\@gobbletwo
    \def\@oddhead{}
    \def\@oddfoot{\hbox to 7 cm{\tiny \versionno
       \hfil}\hskip -7cm\hfil\rm\thepage \hfil {\tiny\draftdate}}
    \def\@evenhead{}\let\@evenfoot\@oddfoot}
\def\draftdate{\number\month/\number\day/\number\year\ \ \ \hourmin }

\long\def\query#1{\hskip 0pt{\vadjust{\everypar={}\small\vtop to 0pt{\hbox{}%
     \vskip -13pt\rlap{\hbox to 49.0pc{\hfil{\vtop{\hsize=8pc\tolerance=6000%
     \hfuzz=.5pc\rightskip=0pt plus 3em\noindent#1}}}}\vss}}}}%
\newcommand\version[1] {\typeout{}\typeout{#1}\typeout{}
                   \vskip3mm \centerline{\fbox{{\tt DRAFT -- #1 -- }
                   {\small\draftdate}}} \vskip3mm}

\def\start{ 
                 \setlength{\textwidth}{17cm} \setlength{\textheight}{25.2cm}
 \hoffset -20mm \topmargin= -13mm \begin{document}\version\versionno
}
\usepackage{amssymb,amsfonts,amsmath,amscd,latexsym,epsf}    

\def \ignore#1 { {} } 

\def \Fig#1#2#3 {
\begin{figure}
\begin{center}
\scalebox{#2}{\includegraphics{#1.eps}}
\label{#1}
\end{center}
\caption{#3}
\end{figure}
}

\def \bea {\begin{eqnarray}}
\def \eea {\end{eqnarray}}
\def \bee {\begin{eqnarray*}}
\def \eee {\end{eqnarray*}}
\def \nn  {\nonumber}
\def \sst {\scriptscriptstyle}
\def \begm {\begin{multline}}
\def \endm  {\end{multline}}
\def \bega {\begin{align}}
\def \enda  {\end{align}}

\def \sp   { \ \ \ , \ \ \ }

\def \tsum  { {\textstyle \sum} }

\def \deg {{-\frac{1}{2b}}}

\def \p {\partial}
\def \pp#1 {{\frac{\p}{\p #1}}}
\def \ppd#1 {{\frac{\p^2}{\p #1 ^2}} }

\def \om {\omega}
\def \g {\gamma}
\def \vf {\varphi}
\def \s {\sigma}
\def \G {\Gamma}  
\def \e {\epsilon}
\def \a {\alpha}
\def \up {\Upsilon_b}
\def \t {\theta}
\def \vt {\vartheta_1}

\def \rar {\rightarrow}
\def \lrar {\leftrightarrow}

\def \bg {\bar{\gamma}}
\def \ap {\alpha'}
\def \ba {\bar{alpha}}
\def \bz {\bar{z}}
\def \bw {\bar{w}}
\def \bx {\bar{x}}
\def \bp {\bar{\partial}}
\def \bq {\bar{q}}
\def \bL {\bar{L}}
\def \bJ {\bar{J}}
\def \tJ {\tilde{J}}
\def \tq {\tilde{q}}
\def \tth {\tilde{\theta}}
\def \ttau {\tilde{\tau}}

\def \Tr {{\rm Tr }\ }
\def \id {{\rm id } }
\def \sgn{ {\rm sgn} }
\def \cst {{\rm constant}}
\def \Res {{\rm Res }}

\def \Z {\mathbb{Z}}
\def \N {\mathbb{N}}
\def \R {\mathbb{R}}
\def \C {\mathbb{C}}
\def \S {\mathbb{S}}

\def \la {\left\langle}
\def \ra {\right\rangle}
\def \lla {\langle\langle}
\def \rra {\rangle\rangle}

\def \uq  {U_q(sl_2)}
\def \asl {\widehat{s\ell_2}}
\def \SLC    {SL(2,\C)}
\def \SLR    {SL(2,\R)}
\def \SLU    {SL(2,\R)/U(1)}
\def \H      {H_3^+}

\def \gsin#1#2 {\left[\!\! \begin{array}{c} #1 \\ #2 \end{array}\!\! \right]}

\def \six#1#2#3#4#5#6{
 _{#1\, #2}\left[\begin{array}{cc} 
#5 & #4 \\ 
#6 & #3
\end{array}\right]}

\def \dft#1#2#3#4#5#6{ \left(\begin{array}{ccc} #1 & #2 & #3 \\ #4 & #5
    & #6 \end{array} \right) }

\def \F32sep { \ , \ }
\def \F32#1#2#3#4#5#6{{}_3F_2 \left(\left.\begin{array}{c}#1 \F32sep #2 \F32sep
    #3 \\ #4\F32sep #5
    \end{array} \right| #6 \right) }

\def\vect#1#2{ \left( \begin{array}{c} #1 \\ #2  \end{array} \right)}

\newcommand {\slidetitle}[1]{ \large \begin{center} { #1} \\
  \end{center} \normalsize \vspace{0.8cm}  }

\usepackage{graphics}\usepackage{epic}
\usepackage{pstricks,pst-coil,pst-node}

\def \mycoil#1#2{\pscoil[coilarm=0,linewidth=.05,coilaspect=0,coilwidth=.4,
coilheight=1.2](2.2,0) \rput[#1]{*0}(2.5,0){$#2$}}

\def \mypoles#1{\psdots[dotstyle=o,dotscale=1.5](0,0)
\psline[linestyle=dashed,dash=3pt 2pt](.15,0)(#1,0)}

\def \mycut#1{\psdots[dotstyle=o,dotscale=1.5](0,0)
\pscoil[coilarm=0,linewidth=.03,coilaspect=0,coilwidth=.4, 
coilheight=1.2](.15,0)(#1,0) }

\newlength{\floatlength}

\def \extline#1#2{\psline[linewidth=.1](2.2,0)\rput[#1]{*0}(2.5,0){$#2$}}

\def \intline#1#2#3{\pcline[linewidth=.1](0,0)(#3,0)
\pssetlength{\floatlength}{#3} \pssetlength{\floatlength}{.5\floatlength}
\rput[#1]{*0}(\floatlength,0.3){$#2$}}

\def \mybdy#1#2#3{\pcline[linewidth=.22](0,0)(#3,0)
\pssetlength{\floatlength}{#3} \pssetlength{\floatlength}{.5\floatlength}
\rput[#1]{*0}(\floatlength,0.3){$\scriptstyle #2$}}

\def\bdy#1{\pspolygon[linecolor=white,fillstyle=hlines,hatchwidth=.5pt,hatchsep=2pt]
(0,0)(#1,0)(#1,-.3)(0,-.3) 
\pcline(0,0)(#1,0)}

\def \degfield#1#2{\psdots[dotscale=1.5,dotstyle=*](0,0)
\rput[#1]{*0}(0.3,0.3){$#2$}}

\def \genfield#1#2{\psdots[dotscale=1.8,dotstyle=+,dotangle=45](0,0)
\rput[#1]{*0}(0.3,0.3){$#2$}}

\def \blockverl#1#2#3#4#5#6#7{
\rput[l]{20}(0,2.5){\extline{l}{#1}}
\rput[l]{160}(0,2.5){\extline{r}{#2}}
\rput[l]{-20}(0,-2.5){\extline{l}{#3}}
\rput[l]{-160}(0,-2.5){\extline{r}{#4}}
\rput[l]{180}(2.2,1.5){\mycoil{r}{\scriptstyle #5}}
\rput[l]{180}(2.2,-1.5){\mycoil{r}{\scriptstyle #6}}
\rput[l]{90}(0,-2.5){\intline{r}{#7}{5}}
}

\def \blockhorl#1#2#3#4#5#6#7{
\rput[l]{70}(2.5,0){\extline{l}{#3}}
\rput[l]{-70}(2.5,0){\extline{l}{#4}}
\rput[l]{110}(-2.5,0){\extline{r}{#1}}
\rput[l]{-110}(-2.5,0){\extline{r}{#2}}
\rput[l]{-90}(1.5,2.2){\mycoil{t}{\scriptstyle #6}}
\rput[l]{-90}(-1.5,2.2){\mycoil{t}{\scriptstyle #5}}
\rput[l]{180}(2.5,0){\intline{t}{#7}{5}}
}

\def \blockhorh#1#2#3#4#5#6#7{
\rput[l]{70}(2.5,0){\extline{l}{#4}}
\rput[l]{-70}(2.5,0){\extline{l}{#3}}
\rput[l]{110}(-2.5,0){\extline{r}{#1}}
\rput[l]{-110}(-2.5,0){\extline{r}{#2}}
\rput[br]{0}(2.5,.3){$\scriptstyle #6$}
\rput[bl]{0}(-2.5,.3){$\scriptstyle #5$}
\rput[l]{180}(2.5,0){\intline{t}{#7}{5}}
}

\def \blockverh#1#2#3#4#5{
\rput[l]{20}(0,2.5){\extline{l}{#4}}
\rput[l]{160}(0,2.5){\extline{r}{#1}}
\rput[l]{-20}(0,-2.5){\extline{l}{#3}}
\rput[l]{-160}(0,-2.5){\extline{r}{#2}}
\rput[l]{90}(0,-2.5){\intline{r}{#5}{5}}
}

\def \vsp {\vspace{3mm}}

\title{Discrete D-branes in $AdS_3$ and in the 2d black hole}

\author{ Sylvain Ribault
\vspace{5mm}
\\
Deutsches Elektronen-Synchrotron \\
Theory Division \\
Notkestrasse 85, Lab 2a \\
Hamburg 22603 \\
Germany  \\
\\ {\tt sylvain.ribault@desy.de }
}

\abstract{
I show how the $AdS_2$ D-branes in the Euclidean $AdS_3$ string theory are related to
the continuous D-branes in Liouville theory. I then propose new discrete
D-branes in the Euclidean $AdS_3$ which correspond to the discrete
D-branes in Liouville theory. These new D-branes satisfy the
appropriate shift equations. They give rise to two families of discrete D-branes
in the 2d black hole, which preserve different symmetries.
}

\preprint{
\hepth{0512238}\\
DESY-05-257
 }

\begin{document}

\section{Introduction and overview}

Liouville theory  and string theories with an affine $\asl$
symmetry have played an important r\^ole in recent studies of
time-dependent string theory, two-dimensional quantum gravity, and the
$AdS$/CFT correspondence.
The features of these theories which are
well-understood suggest that they share many important
properties. This is not surprising considering that the Virasoro
algebra of Liouville theory can be obtained from the $\asl$ affine Lie
algebra by a quantum Hamiltonian reduction \cite{bffow89}. This
suggests that Liouville theory can be found as a subsector of theories
with an $\asl$ symmetry. For example, $AdS_3$
string theory can be reduced to Liouville theory via a topological
twist \cite{mv93,rw05}.

Conversely, it would be interesting to reconstruct the full
$AdS_3$ string theory in terms of the better-understood Liouville
theory. A hint that this can be done comes from Zamolodchikov and
Fateev's relation \cite{fz86} between the Knizhnik--Zamolodchikov (KZ)
and Belavin--Polyakov--Zamolodchikov (BPZ) systems of differential
equations, which reflect the $\asl$ and Virasoro symmetries
respectively. More recently, all correlation functions of the $\H$
model (the Euclidean version of $AdS_3$ string theory) on a sphere
have been written in terms of Liouville correlation functions
\cite{rt05}. Proving this relation relied on the prior knowledge of
these $\H$ correlation functions in terms of well-characterized
objects, namely the three-point structure constants and the conformal
blocks. However, the $\H$-Liouville relation would be most useful if
it allowed the construction of previously unknown objects in the $\H$
model from known objects in Liouville theory.
One purpose of this article is to demonstrate that it indeed does.

The new objects in the $\H$ model which I plan to construct are discrete
D-branes (in both meanings of having a discrete open string spectrum
and coming in a discrete family) 
which correspond to the Zamolodchikov--Zamolodchikov (ZZ) D-branes
in Liouville theory \cite{zz01}. I will first determine a relation between the known
continuous $AdS_2$ branes in the $\H$ model \cite{pst01} and the
continuous Fateev--Zamolodchikov--Zamolodchikov--Teschner (FZZT) D-branes in
Liouville theory \cite{fzz00,tes00}. The main feature of this relation
is the correspondence (\ref{spm}) between the parameters of these families of
D-branes, which associates two different FZZT branes to one $AdS_2$
brane. Moreover, it is possible to relate the correlators of bulk fields in the
presence of FZZT and $AdS_2$ branes eq. (\ref{prvs}), but only in a
particular regime which I will call the bulk regime. This is due to
singularities in the $\H$ conformal blocks, which have a clear
interpretation -- but so far no resolution -- in terms of Liouville
theory.

The relation between FZZT and $AdS_2$ branes will then suggest a
natural ansatz for a family of discrete branes in $\H$ parametrized by
two integers $(m,n)$, related to the ZZ branes of
Liouville theory. The most useful characterization of these branes,
which I will call $AdS_2^d$ branes, 
is the relation to the $AdS_2$
branes eq. (\ref{aad}). (The name $AdS_2^d$ refers to that relation
and not to the geometry of the new discrete branes.)
These $AdS_2^d$ branes will be shown to be solutions of the same shift
equation that was checked for the $AdS_2$ branes \cite{pst01}. How to
modify this equation for the case of discrete branes will be suggested by
Liouville theory. I will then propose a tentative relation between
$\asl$ representations and D-branes in $\H$, inspired by the Cardy
relation which holds in rational conformal field theories, and which
may help understand
which $\H$ D-branes can be related to Liouville branes and which ones
cannot.

From the new discrete D-branes in the $\H$ model, two families of
compact D-branes
in the
2d ``cigar'' black hole $\SLU$ obeying two different gluing conditions
can be constructed along the lines of
\cite{rs03}. Some of these D-branes have a geometric interpretation
as D0-branes at the tip of the cigar, the others do not have any
geometric interpretation. These new D-branes in the 2d black hole
 can then easily be translated into
D-branes in the $N=2$ Liouville theory in Hosomichi's formalism
\cite{hos04}, which provides a second independent
shift equation.


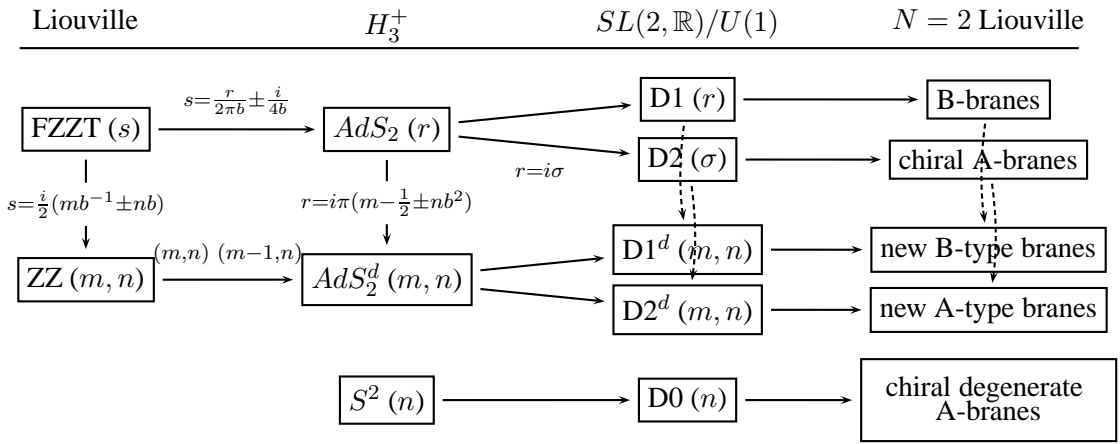
\begin{center}
\psset{unit=.4cm}
\pspicture[](1.5,0)(30,14)
\rput[t](0,14){Liouville}
\rput[t](10,14){$\H$}
\rput[t](20,14){$\SLU$}
\rput[t](30,14){$N=2$ Liouville}
\psline(-2.2,12.7)(33,12.7)

\rput(0,10){\rnode{FZZT}{\psframebox{FZZT ($s$)}}}
\rput(0,5){\rnode{ZZ}{\psframebox{ZZ ($m,n$)}}}

\rput(10,10){\rnode{AdS2}{\psframebox{$AdS_2$ ($r$)}}}
\rput(10,5){\rnode{AdS2d}{\psframebox{$AdS_2^d$ ($m,n$)}}}

\rput(20,11){\rnode{D1}{\psframebox{D1 ($r$)}}}
\rput(20,9){\rnode{D2}{\psframebox{D2 ($\sigma$)}}}
\rput(20,6){\rnode{D1d}{\psframebox{D1$^d$ ($m,n$)}}}
\rput(20,4){\rnode{D2d}{\psframebox{D2$^d$ ($m,n$)}}}

\rput(30,11){\rnode{Bb}{\psframebox{B-branes}}}
\rput(30,9){\rnode{Ab}{\psframebox{chiral A-branes}}}
\rput(30,6){\rnode{nBb}{\psframebox{new B-type branes}}}
\rput(30,4){\rnode{nAb}{\psframebox{new A-type branes}}}

\rput(10,1){\rnode{S2}{\psframebox{$S^2$ ($n$)}}}
\rput(20,1){\rnode{D0}{\psframebox{D0 ($n$)}}}
\rput(30,1){\rnode{dAb}{\psframebox{\begin{tabular}{c} chiral
	  degenerate \vspace{-2mm} \\ A-branes \end{tabular}}}}
\ncline[nodesep=3pt]{->}{S2}{D0}
\ncline[nodesep=3pt]{->}{D0}{dAb}

\ncline[nodesep=3pt]{->}{FZZT}{ZZ}
\lput*{0}{$\scriptstyle s=\frac{i}{2}(mb^{-1}\pm nb)$}
\ncline[nodesep=3pt]{->}{AdS2}{AdS2d}
\lput*{0}{$\scriptstyle r=i\pi(m-\frac12\pm nb^2)$}
\ncarc[nodesep=1.5pt,linestyle=dashed, dash=2pt 1.4pt]{->}{D2}{D2d}
\ncarc[nodesep=1.5pt,linestyle=dashed, dash=2pt 1.4pt]{<-}{D1d}{D1}
\ncarc[nodesep=1.5pt,linestyle=dashed, dash=2pt 1.4pt]{->}{Ab}{nAb}
\ncarc[nodesep=1.5pt,linestyle=dashed, dash=2pt 1.4pt]{<-}{nBb}{Bb}
\ncline[nodesep=3pt]{->}{FZZT}{AdS2}
\aput{0}{$\scriptstyle s=\frac{r}{2\pi b}\pm \frac{i}{4b}$}
\ncline[nodesep=3pt]{->}{ZZ}{AdS2d}
\aput{0}{$\scriptstyle (m,n)\ (m-1,n)$}
\ncline[nodesep=3pt]{->}{AdS2}{D1}
\ncline[nodesep=3pt]{->}{AdS2}{D2}
\bput{0}{$\scriptstyle r=i\sigma$}
\ncline[nodesep=3pt]{->}{AdS2d}{D1d}
\ncline[nodesep=3pt]{->}{AdS2d}{D2d}
\ncline[nodesep=3pt]{->}{D1}{Bb}
\ncline[nodesep=3pt]{->}{D2}{Ab}
\ncline[nodesep=3pt]{->}{D1d}{nBb}
\ncline[nodesep=3pt]{->}{D2d}{nAb}

\endpspicture
\end{center}

\section{$AdS_2$ D-branes from Liouville theory}

The aim of this section is to generalize the relation between $\H$ and
Liouville bulk correlators on the Riemann sphere \cite{rt05} to correlators of bulk fields
in the presence of worldsheet boundaries described by
continuous D-branes: the $AdS_2$ branes
on the $\H$ side, the FZZT branes on the Liouville side.

The relation between bulk correlators on the sphere can be decomposed into relations
between bulk conformal blocks on the one hand, and bulk three-point
structure constants on the other hand \cite{rib05}. The introduction
of a worldsheet boundary implies a modification of the conformal
blocks, and the introduction of extra structure constants, namely the
one-point functions (which must vanish when no boundary is there to break the
worldsheet translation invariance). 

Let me briefly recall that Liouville theory is a two-dimensional
conformal field theory on a worldsheet parametrized by a complex
number $z$. The theory may be defined in terms of a field $\phi(z)$ by
the action:
\bea
S^{\rm Liouville}=\int d^2z\ \left(|\p_z\phi|^2+\mu_L e^{2b\phi}\right)\ .
\eea
The $\H$ model describes strings in a three-dimensional space and
therefore requires three fields $\phi,\gamma,\bar{\gamma}$:
\bea
S^{\H}=k\int d^2z\ \left(|\p_z\phi|^2 +e^{2\phi} \p \g \bp\bg \right) \ .
\eea
A more complete review with relevant references can be found in
\cite{rt05}. 

\subsection{Comparison of one-point functions}

Consider one-point functions 
 of the closed string worldsheet fields $V_\a(z)$ in
Liouville theory and $\Phi^j(x|z)$ in the $\H$ model. From the bulk
$\H$-Liouville relation, the Liouville momentum $\a$ and the
$\H$ spin $j$ are related by:
\bea
\a=b(j+1)+\frac{1}{2b},
\label{aj}
\eea
where the Liouville parameter $b$ is related to the $\H$ model level
$k$ by $b^2=\frac{1}{k-2}$. 
In terms of $j$, the one-point function for the Liouville
FZZT brane parametrized by the real number $s$ is \cite{fzz00,tes00}
\begin{multline}
\la V_{\a=b(j+1)+\frac{1}{2b}}(z)\ra_s
=\frac{\Psi^{\rm FZZT}_s}{|z-\bar{z}|^{2\Delta_\a}} ,
\\
\Psi^{\rm FZZT}_s=
(\pi \mu_L
\gamma(b^2))^{-j-\frac12}\frac{1}{\pi 2^\frac14 b}
\Gamma(2j+1) \Gamma(1+b^2(2j+1)) \cosh 2\pi bs (2j+1) ,
\label{pfzzt}
\end{multline}
where $z$ is the complex worldsheet coordinate and $\mu_L$ the
Liouville interaction strength. The one-point function for an
$AdS_2$ brane in $\H$ with real parameter $r$ is: \cite{pst01,lpo01}
\bea
\Psi^{AdS_2}_r(x) = \nu_b^{j+\frac12}
(8b^2)^{-\frac14 }
|x+\bar{x}|^{2j}\Gamma(1+b^2(2j+1)) e^{-r(2j+1) \sgn
(x+\bar{x})} ,
\label{pxads}
\eea
where $\nu_b=\pi\frac{\G(1-b^2)}{\G(1+b^2)}$ so that $\Phi^{j=0}$
is the identity field (in slight contrast to \cite{pst01}), and $x$ is
a complex isospin variable which labels states within a continuous $\SLC$
representation of spin $j$. 

This one-point function of the $x$-basis fields is
written here for later use, but is not clearly related to the one-point
function in Liouville theory. Instead, the bulk $\H$-Liouville
relation suggests to consider the $\mu$-basis fields
\bea
\Phi^j(\mu|z)=|\mu|^{2j+2}\int_\C d^2x\ e^{\mu x-\bar{\mu}\bx}
\Phi^j(x|z),
\label{pmpx}
\eea
whose one-point functions are obtained from eq. (\ref{pxads}) 
after a straightforward calculation:
\begin{multline}
\la \Phi^j(\mu|z)\ra_r=
\frac{ \Psi^{AdS_2}_r}{|z-\bar{z}|^{2\Delta_j}},
\\
\Psi^{AdS_2}_r =|\mu|\delta(\Re \mu) \nu_b^{j+\frac12} 
 \pi (8b^2)^{-\frac14 }
\Gamma(2j+1)\Gamma(1+b^2(2j+1))  \cosh(2j+1)(r-i\frac{\pi}{2} \sgn \Im \mu).
\label{pmads}
\end{multline}
It is now obvious that the $AdS_2$ D-brane one-point function is essentially the
same as that of an FZZT brane (\ref{pfzzt}), but depending on $\sgn
\Im \mu$ two different boundary parameters may appear:
\bea
s_\pm = \frac{r}{2\pi b} \pm \frac{i}{4b}.
\label{spm}
\eea
Such a relation could have been expected on several grounds. First,
the FZZT branes are invariant under $s\rar -s$ whereas the $AdS_2$ branes
are not invariant under $r\rar -r$, so there cannot be a one-to-one
relation between the parameters $r$ and $s$. Second, the $\SLR$
symmetry of the $AdS_2$ brane, which acts on the $x$ parameter,
 does not completely determine the $x$
dependence of the one-point function, but allows an arbitrary
dependence on $\sgn(x+\bx)$ \cite{pst01}. Therefore the
one-point function for an $AdS_2$ brane involves two structure
constants (instead of one in Liouville theory), which in the $\mu$
basis are encoded in the $\sgn \Im \mu$ dependence. Third, the
difference $s_+-s_-=\frac{i}{2b}$ is the jump in Liouville boundary
condition induced by a boundary degenerate field
$B_{-\frac{1}{2b}}$. This is not surprising in view of the appearance
  of such degenerate fields in the $\H$-Liouville relation beyond the
  one-point function discussed below.

\subsection{Comparison of conformal blocks}

The $\H$ bulk 
conformal blocks are controlled by the Knizhnik--Zamolodchikov
equations \cite{kz84}, which are enough to determine their relation
with Liouville conformal blocks \cite{rib05}. Let me determine the KZ
equations satisfied by the conformal blocks involved in the correlator
of $n$ bulk fields in the presence of an $AdS_2$ brane
$\la \Phi^{j_1}(\mu_1|z_1)\cdots
\Phi^{j_n}(\mu_n|z_n)\ra_r $.
In Wess--Zumino--Witten models with  
symmetry-preserving boundary conditions, such KZ equations are
identical to the KZ equations satisfied by a correlator of $2n$ bulk fields on
the sphere (at points $z_1,\cdots z_n,\ \bz_1\cdots \bz_n$), 
modulo a twist of the currents acting on the
reflected fields if the gluing conditions are non-trivial. In the case
of $AdS_2$ branes, the gluing conditions are trivial as I will now
show. 

Let me call $J^a(z),\bJ^a(\bz)$ the left- and right-moving currents
of the $\H$ model \cite{gaw91}. Their modes generate an
$\asl(\C)\times \asl(\C)$ affine Lie algebra. Their zero modes act on
the fields $\Phi^j(x|z)$ or $\Phi^j(\mu|z)$ as differential operators
with respect to the isospin variables $x$ or $\mu$:
\bea
\begin{array}{lll} J^-_0 & = \pp{x} & = \mu \ ,
  \\
J^0_0 & = x\pp{x} -j & = -\mu\pp{\mu} \ ,
 \\
J^+_0 & = x^2\pp{x} -2jx & = \mu \ppd{\mu} -\frac{j(j+1)}{\mu} \ ,
\end{array}
\eea 
and the currents $\bJ^a_0$ are defined by repacing $x,\mu$ with
$\bar{x},\bar{\mu}$. Note that this definition of the $\bJ^a_0$
currents is incompatible with the change of basis (\ref{pmpx}) and is
therefore basis-dependent. As a result, the gluing conditions will
also be basis-dependent.

The $\mu$-basis one-point function of the $AdS_2$ brane satisfies
$(J^a_0+\bJ^a_0)\Psi^{AdS_2}_r(\mu)=0$, which corresponds to the trivial gluing condition
$J=\bJ$ (see for instance \cite{sch02}). Thus, it satisfies the same
KZ equations as the bulk two-point function
$\la\Phi^{j}(\mu|z)\Phi^j(\bar{\mu}|\bz)\ra$. Indeed, the $\mu$-dependences
are similar: $|\mu|^2\delta^{(2)}(\mu+\bar{\mu})$ for the bulk
two-point function, $\mu \delta(\mu+\bar{\mu})$ for the one-point
function.
In contrast, the $x$-basis one-point function has an
$|x+\bar{x}|^{2j}$ factor which contrasts with the bulk two-point function
$|x-\bar{x}|^{4j}$. This reflects the fact that the gluing conditions are
non-trivial
 in the $x$-basis. 

Since the correlator $\la \Phi^{j_1}(\mu_1|z_1)\cdots
\Phi^{j_n}(\mu_n|z_n)\ra_r $ satisfies the same KZ equations as a bulk
correlator with $2n$ fields, these equations are equivalent to BPZ
equations via Sklyanin's separation of variables, as explained in
\cite{rt05,sto00}.
This leads to the following relation between $\H$
and Liouville correlators in the presence of worldsheet boundaries,
where the equality so far means ``satisfies the same differential
equations as'':
\begin{multline}
\la \prod_{\ell=1}^n \Phi^{j_\ell}(\mu_\ell|z_\ell) \ra_r = \pi^2
\sqrt{\frac{b}{2}} (-1)^n  \left|{\scriptstyle \sum_{i=1}^n}\Re (\mu_i z_i)
\right| \delta\left(\Re({\scriptstyle
  \sum_{i=1}^n}\mu_i)\right)     
\\
\times 
|\Theta_{2n}|^{\frac{k-2}{2}} \la
\prod_{\ell=1}^n V_{\a_\ell}(z_\ell)\prod_{a=1}^{n-1}
V_{-\frac{1}{2b}}(y_a) \ra_{s=\frac{r}{2\pi b} - \frac{i}{4b}\sgn\sum_{i=1}^n\Im
  \mu_i}, 
\label{prvs}
\end{multline}
In this equation the following conventions are used:
the momenta and spins are related as in eq. (\ref{aj}), I assume
$\mu_L=\frac{b^2}{\pi^2}$, the function $\Theta_{2n}$ is defined by
\bea 
\Theta_{2n}=\frac{\prod_{\ell< \ell'\leq n}|z_{\ell\ell'}|^2 \prod_{\ell,
  \ell'\leq n} (z_\ell-\bz_{\ell'}) \prod_{a<a'\leq n-1} |y_{aa'}|^2
\prod_{a,a'\leq n-1} (y_a-\bar{y}_{a'})}{
\prod_{\ell=1}^n\prod_{a=1}^{n-1}
|z_\ell-y_a|^2 |z_\ell-\bar{y}_a|^2} , 
\eea
and most importantly the
$y_a$ are the roots with positive imaginary parts of the real
polynomial $P(t)$ defined by:
\bea
\sum_{\ell=1}^n \left( \frac{\mu_\ell}{t-z_\ell}
+\frac{\bar{\mu_\ell}}{t-\bz_\ell} \right) = \left[{
\sum_{\ell=1}^n} (\mu_\ell z_\ell+\bar{\mu}_\ell \bz_\ell) \right]
\frac{P(t)}{\prod_{\ell=1}^n (t-z_\ell)(t-\bz_\ell)} .
\label{defp}
\eea

In the case $n=3$, the equation (\ref{prvs}) can be represented as:
\bea
\psset{unit=.4cm}
  \pspicture[](-5,-3.9)(7,2.4)
    \rput[l]{0}(-5,0){\bdy{10}}
    \rput[l]{-90}(-4,2){\genfield{l}{z_1}}
    \rput[l]{-90}(-1,2){\genfield{l}{z_2}}
    \rput[l]{-90}(4,2){\genfield{l}{z_3}}
    \rput[l](5.5,0){$r$}
   \rput[b](0,-4){$\H$ model}
{ \gray
    \rput[l]{-90}(-4,-2){\genfield{l}{\bz_1}}
    \rput[l]{-90}(-1,-2){\genfield{l}{\bz_2}}
    \rput[l]{-90}(4,-2){\genfield{l}{\bz_3}} 
}
\endpspicture 
\propto
\psset{unit=.4cm}
  \pspicture[](-7,-3.9)(9,2.4)
    \rput[l]{0}(-5,0){\bdy{10}}
    \rput[l]{-90}(-4,2){\genfield{l}{z_1}}
    \rput[l]{-90}(-1,2){\genfield{l}{z_2}}
    \rput[l]{-90}(4,2){\genfield{l}{z_3}}
    \rput[l]{-90}(-2,1){\degfield{l}{y_1}}
    \rput[l]{-90}(1,1.5){\degfield{l}{y_2}}
\rput[l](5.5,0){${\scriptstyle s=\frac{r}{2\pi b} \pm \frac{i}{4b}
  }$}
    \rput[b](0,-4){Liouville theory}
 {\gray 
    \rput[l]{-90}(-4,-2){\genfield{l}{\bz_1}}
    \rput[l]{-90}(-1,-2){\genfield{l}{\bz_2}}
    \rput[l]{-90}(4,-2){\genfield{l}{\bz_3}}
\rput[l]{-90}(-2,-1){\degfield{l}{\bar{y}_1}}
    \rput[l]{-90}(1,-1.5){\degfield{l}{\bar{y}_2}}
 }
\endpspicture
\eea
The reflected fields at $(\bz_1\cdots \bz_n)$ in the lower half-plane are not physical, but
they are indicated in this picture because they appear in the KZ or
BPZ equations satisfied by the physical correlators of eq. (\ref{prvs}).

In this subsection I only argued  that both sides of equation (\ref{prvs}) satisfy
identical systems of differential equations. This amounts to a relation between
the conformal blocks from which the correlators are built.
In the next subsection I will complete the argument for equation
(\ref{prvs}) and show that it holds in a certain regime.

\subsection{The bulk regime}

From the explicit expressions for the one-point functions
(\ref{pfzzt}), (\ref{pmads}) it is easy to check that the equation
(\ref{prvs}) holds in the case $n=1$, which does not involve any
insertion of degenerate Liouville fields $V_{-\frac{1}{2b}}$. One
could then think that it is possible to prove equation (\ref{prvs}) by
a recursion on $n$, using the bulk operator product expansion to
reduce the case of the $n$-point function in the limit $z_1\rar z_2$ to the case of the
$n-1$-point function. (The bulk OPEs in the
$\H$ model and Liouville theory are indeed 
related in a way which would suit such an argument \cite{rt05}.) Then
one would rely on the KZ equation to extend the relation (\ref{prvs})
to all values of $z_i$, away from the limit $z_1\rar z_2$. 

However,
this argument does not work because the conformal blocks which solve
the KZ equations
have singularities. These
singularities are most easily seen in the corresponding 
Liouville theory conformal blocks: they occur whenever one of the $y_a$ becomes
real. Indeed the $y_a$ are defined as the roots of the real polynomial
$P(t)$ (\ref{defp}). Such a polynomial can have real roots and pairs
of complex conjugate roots. Let me call the {\it bulk
  regime} the range of values of $\mu_\ell, z_\ell$ such that all
the roots of $P(t)$ are complex. The repeated use of the bulk OPE
$z_1\rar z_2\rar\cdots z_n$ (as required by the recursion above) 
is possible only in the bulk regime,
because the definition of $P(t)$
(\ref{defp}) implies that for $z_1\rar z_2$ some root $y_1$ of $P(t)$
will also move close to $z_1$, and therefore in the bulk. 
Thus, the
equation (\ref{prvs}) holds only in the bulk regime. 
Unfortunately, this prevents the easy
determination of an $\H$-Liouville relation in other bases like the
$x$ basis, which
would involve an integration over all values of $\mu_\ell$. 

Let me illustrate the singularities of the conformal blocks in the
case of a two-point function 
$\la \Phi^{j_1}(\mu_1|z_1) \Phi^{j_2}(\mu_2|z_2) \ra_r$. In this case
the polynomial $P(t)$ has degree two and its roots
are complex provided 
\bea
z\equiv \left|\frac{z_1-\bz_2}{z_1-z_2}\right| >
\frac{|\mu_1|+|\mu_2|}{|\mu_1+\mu_2|} \ .
\eea
The cross-ratio $z$ varies from $1$ when the two $\H$ bulk fields are
far apart or close to the boundary, to $+\infty$ when they are close
together or far from the boundary. The corresponding Liouville
configurations are:
\bea
\begin{array}{c}
\psset{unit=.4cm}
\pspicture[](-3,-3.9)(5,2.4)
    \rput[l]{0}(-3.5,0){\bdy{7}}
    \rput[l]{-90}(-2.9,1.1){\genfield{l}{z_1}}
    \rput[l]{-90}(2.9,1.1){\genfield{l}{z_2}}
        \rput[l]{45}(-1,0){\degfield{l}{y_1}}
 \rput[l]{45}(1,0){\degfield{l}{y_2}}
       \rput[b](0,-4.5){Boundary regime}
 {\gray 
    \rput[l]{-90}(-2.9,-1.1){\genfield{l}{\bz_1}}
    \rput[l]{-90}(2.9,-1.1){\genfield{l}{\bz_2}}
    }
\endpspicture
\pspicture[](-5,-3.9)(5,2.4)
    \rput[l]{0}(-3.5,0){\bdy{7}}
    \rput[l]{-90}(-2.5,2){\genfield{l}{z_1}}
    \rput[l]{-90}(2.5,2){\genfield{l}{z_2}}
        \rput[l]{-90}(0,0){\degfield{l}{}}
       \rput[b](0,-4.5){Singularity}
 {\gray 
    \rput[l]{-90}(-2.5,-2){\genfield{l}{\bz_1}}
    \rput[l]{-90}(2.5,-2){\genfield{l}{\bz_2}}
    }
\endpspicture
\pspicture[](-5,-3.9)(8,2.4)
    \rput[l]{0}(-3.5,0){\bdy{7}}
    \rput[l]{-90}(-2,2.6){\genfield{l}{z_1}}
    \rput[l]{-90}(2,2.6){\genfield{l}{z_2}}
        \rput[l]{-90}(0,1){\degfield{l}{y}}
       \rput[b](0,-4.5){Bulk regime}
 {\gray 
    \rput[l]{0}(-2,-2.6){\genfield{l}{\bz_1}}
    \rput[l]{0}(2,-2.6){\genfield{l}{\bz_2}}
    \rput[l]{-90}(0,-1){\degfield{l}{\bar{y}}}
 }
\endpspicture
\\
\psset{unit=.4cm}
\pspicture[](-13,-2)(18,1)
\psline[linewidth=1.6pt]{->}(-15,0)(15,0)
\rput[t](0,-.5){$\frac{|\mu_1|+|\mu_2|}{|\mu_1+\mu_2|}$}
\rput[t](-15,-.5){$1$}
\rput[t](14.5,-.5){$+\infty$}
\rput[l](16,0){$z=\left|\frac{z_1-\bz_2}{z_1-z_2}\right|$}
\psdots[dotstyle=|,dotscale=1.2](-15,0)(0,0)
\endpspicture
\end{array}
\label{bsb}
\eea
In the boundary regime, the relation between the KZ and BPZ equations
still holds. However it is not clear that a relation between $\H$ and
Liouville correlators can be found. Such a relation would have to
specify which boundary parameters appear in Liouville theory. 
The boundary degenerate fields induce jumps of the boundary parameter $s$
by the quantity $\frac{i}{2b}$ \cite{fzz00}. The fact that the two
boundary parameters $s_\pm$ (\ref{spm}) differ by this quantity is
very suggestive, but more work needs to be done. This issue is however
not relevant to the present article, whose purpose is to find new
discrete D-branes in the $\H$ model.

\section{More branes in the Euclidean $AdS_3$}

In this section I will show that the relation between Liouville FZZT branes
and $AdS_2$ branes in the $\H$ model suggests a natural ansatz for new
discrete D-branes in the $\H$ model, which will be related to the
discrete ZZ-branes in Liouville theory. This ansatz will then be
subjected to a number of tests.

Let
me first briefly review the ZZ branes and their relation to the
continuous FZZT branes. The ZZ branes are parametrized by two strictly
positive integers $(m,n)$ and are described by the one-point functions
\cite{zz01} 
\begin{multline}
\la V_{\a=b(j+1)+\frac{1}{2b}}(z)\ra_{(m,n)}
=\frac{\Psi^{\rm ZZ}_{(m,n)}}{|z-\bar{z}|^{2\Delta_\a}} ,
\\
\Psi^{\rm ZZ}_{(m,n)}=
(\pi \mu_L
\gamma(b^2))^{-j-\frac12}\frac{2^\frac34}{\pi  b}
\Gamma(2j+1) \Gamma(1+b^2(2j+1)) \sin \pi m (2j+1) \sin \pi n
b^2(2j+1) \ .
\label{pzz}
\end{multline}
A well-known property of these ZZ branes which will be 
most useful in the following is:
\bea
\Psi^{\rm ZZ}_{(m,n)}=\Psi^{\rm
  FZZT}_{\frac{i}{2}(mb^{-1}+nb)}-\Psi^{\rm FZZT}_{\frac{i}{2}(mb^{-1}-nb)}.
\label{zzfzzt}
\eea

\subsection{An ansatz for new discrete D-branes}

The previous section demonstrated that an $AdS_2$ brane with boundary
parameter $r$  is related
to FZZT branes with boundary parameters $s=\frac{r}{2\pi
  b}-\frac{i}{4b}\sgn \Im \mu$. It is natural to look for
discrete branes in the $\H$ model which would preserve the same
symmetries as the $AdS_2$ branes (in other words, they would obey the
same gluing conditions)
and which would be related in a similar
manner to ZZ branes with parameters depending on $\sgn \Im \mu$. 
In addition, I have explained that the difference of the two possible boundary
parameters has an interpretation as the jump induced by a
boundary degenerate field, which is quite natural considering the
appearance of such fields in the boundary regime (\ref{bsb}). 
Through the relation between ZZ and FZZT
branes eq. (\ref{zzfzzt}), this jump corresponds to a jump $m\rar m-1$
of the parameter $m$ of the ZZ branes. This suggests the following
relation:
\bea
\begin{array}{ccc}
{\rm New\ discrete\ brane\ in\ }\H& \hspace{1.3cm}  & {\rm Liouville\ ZZ\ branes}
\\
(m,n)\ {\rm strictly\ positive\ integers} & &\left\{ \begin{array}{ll} 
(m-1,n)\ & {\rm if}\ \ \sgn\Im\mu > 0
\\ (m,n)\ & {\rm if}\ \ \sgn\Im\mu < 0
\end{array} \right.
\end{array}
\eea 
I will call ``discrete $AdS_2$ branes'' or ``$AdS_2^d$ branes'' these
new branes. 
Their above definition in terms of ZZ branes can
be translated into a relation with $AdS_2$ branes via the ZZ-FZZT
relation eq. (\ref{zzfzzt}) and the $AdS_2$-FZZT relation of the
previous section:
\bea
\Psi^{AdS_2^d}_{(m,n)}=\Psi^{AdS_2}_{i\pi(m-\frac12+nb^2)}
-\Psi^{AdS_2}_{i\pi(m-\frac12-nb^2)} \ .
\label{aad}
\eea
The essential feature of this relation is the shift $-\frac12$, which
directly corresponds to the shift of the boundary parameters in the $AdS_2$-FZZT relation
eq. (\ref{spm}). Let me write explicitly the one-point function of the
$AdS_2^d$ branes in the $x$ basis:
\begin{multline}
\Psi^{AdS_2^d}_{(m,n)}(x) = \nu_b^{j+\frac12}
(8b^2)^{-\frac14 }
|x+\bar{x}|^{2j}\Gamma(1+b^2(2j+1))
\\
\times 2i\sgn(x+\bar{x})\
e^{-i\pi(m-\frac12)(2j+1) \sgn
(x+\bar{x})}  \sin\pi b^2 n(2j+1)\ .
\label{pxadsd}
\end{multline}
Naturally, the relation (\ref{aad}) provides a simple way to derive
the one-point functions of the $AdS_2^d$ branes in any basis. I have
used the $x$ basis because the shift equation of the next subsection
is formulated in this basis.

\subsection{Verification of the shift equation \label{vse} }

The one-point function for an $AdS_2$ brane was found in \cite{pst01}
by solving a shift equation indicating how it should behave under
shifts $j\rar j\pm \frac12$. A modified version of the shift equation
is expected to hold for discrete branes preserving the same
symmetries. This expectation is based on the study of shift equations
for ZZ and FZZT branes in Liouville theory \cite{fzz00,zz01} which I now review. 

The shift equations for ZZ and FZZT branes are of the type:
\bea
R^a_s \Psi^a_s(\a)=F_- \Psi^a_s(\a-\frac{b}{2}) +F_+
\Psi^a_s(\a+\frac{b}{2}) \ ,
\label{rppp}
\eea
where the index $a$ means ZZ or FZZT, with brane parameters
generically called $s$. The coefficients $F_\pm$ on the right-hand
side do not
depend on the type or parameter of the D-brane because they are
fusing matrix elements. However, the quantity $R^a_s$ 
depends on the type of brane: \footnote{In the article
  \cite{zz01} the denominator $\Psi^{\rm ZZ}(\a=0)$ is absent from
  $R^{\rm ZZ}_{(m,n)}$ because the one-point function is normalized
  so that $\Psi^{\rm ZZ}(\a=0)=1$.}
\bea
R^{\rm FZZT}_s &= & R^{\rm
  FZZT}(-\frac{b}{2},Q|s)=-2\pi\sqrt{\frac{\mu_L}{\sin \pi b^2}}
\frac{\G(-1-2b^2)}{\G(-b^2)^2} \cosh 2\pi b s    \ ,
\\
R^{\rm ZZ}_{(m,n)} & =& \frac{\Psi^{\rm ZZ}_{(m,n)}(\a
  =-\frac{b}{2})}{\Psi_{(m,n)}^{\rm ZZ}(\a=0)} \ ,
\eea
where the bulk-boundary structure constant value $R^{\rm
  FZZT}(-\frac{b}{2},Q|s)$ was derived in \cite{fzz00} by a free field
computation and can also be deduced from the general formula for the
bulk-boundary structure constant 
\cite{hos01} by carefully taking the relevant limit as sketched in \cite{sch05}. 

One may wonder how the shift equations (\ref{rppp}), where the factor
$R^a_s$ depends on the type of brane ($a\in\{$ZZ,FZZT$\}$), can be compatible
with the linear relation (\ref{zzfzzt}) between ZZ and FZZT branes. The
compatibility actually requires the non-trivial relation  $R^{\rm
  ZZ}_{(m,n)}  = R^{\rm FZZT}_{s=i\frac{mb^{-1}+nb}{2}} = R^{\rm
  FZZT}_{s=i\frac{mb^{-1}-nb}{2}} $. A direct computation shows that
this relation is indeed obeyed:
\bea
\frac{\Psi^{\rm ZZ}_{(m,n)}(
  -\frac{b}{2})}{\Psi^{\rm ZZ}_{(m,n)}(0)}
 = R^{\rm FZZT}_{i\frac{mb^{-1}+nb}{2}} = R^{\rm
  FZZT}_{i\frac{mb^{-1}-nb}{2}} = -2\pi\sqrt{\frac{\mu_L}{\sin \pi b^2}}
\frac{\G(-1-2b^2)}{\G(-b^2)^2} (-1)^m \cos \pi n b^2 .
\label{rzrf}
\eea

This analysis of the Liouville branes' shift equations can be
generalized to $\H$ branes' shift equations. The continuous $AdS_2$
branes are indeed known to satisfy an equation of the type
\cite{pst01}
\bea
R^{AdS_2}_r \Psi^{AdS_2}_r(j) = F_-^{\H} \Psi^{AdS_2}_r(j-\frac12)
+F_+^{\H} \Psi^{AdS_2}_r(j+\frac12).
\eea
In the notations of \cite{pst01}, the quantity $R^{AdS_2}_r$ can be
computed explicitly as $R^{AdS_2}_r=(x+\bar{x}) B(\frac12) A(\frac12,0|r)$ (see in
particular the equation (3.28) therein) \footnote{Note that $\nu_b=\pi\frac{\G(1-b^2)}{\G(1+b^2)}$
now has an extra factor $\pi$ wrt \cite{pst01}
  so that $\Phi^{j=0}$ is the identity field, see eq. (\ref{pxads}) of
  the present article and footnote 7 of \cite{pst01}. Also note that
  the requirement $B(j=0)=1$ leads to a different sign for $B(j)$ as
  compared to \cite{pst01}.}. 

Now the shift equation for discrete branes in $\H$ should be
identical to that for continuous branes, except for the replacement of
$R^{AdS_2}_r$ with
\bea
R^{AdS_2^d}_{(m,n)}=\frac{\Psi^{AdS_2^d}_{(m,n)}(j=\frac12)}{\Psi^{AdS_2^d}_{(m,n)}(j=0)}.
\eea
Does the ansatz (\ref{aad}) satisfy the resulting shift equation? Like in
Liouville theory, the shift equation for discrete branes boils down to
the equations
\bea
R^{AdS_2^d}_{(m,n)}  \overset{!}{=} R^{AdS_2}_{r=i\pi(m-\frac12 + nb^2)}
\overset{!}{=} R^{AdS_2}_{r=i\pi(m-\frac12 - nb^2)}.
\eea
These equations can now be checked by direct calculation, and the
three quantities to be compared are indeed all equal to
\bea
 2i|x+\bar{x}| \sgn(x+\bar{x})
\sqrt{\nu_b} 
\frac{\G(1+2b^2)}{\G(1+b^2)} (-1)^m
\cos \pi nb^2\ .
\eea

\subsection{Checks and interpretations {\it \`a la Cardy}}

\subsubsection{D-branes and representation theory}

Let me discuss how the proposed discrete $AdS_2$ branes help to complete
 the list of D-branes in the Euclidean $AdS_3$. 
Cardy has shown that in rational
two-dimensional conformal field theories, 
symmetry-preserving D-branes are naturally associated to
representations of the relevant symmetry algebra
\cite{car89}. This idea can be extended to Liouville
 theory. To start with, the continuous FZZT branes are naturally
 associated to the continuous representations of the Virasora algebra,
 which appear in the physical Liouville spectrum and have momenta
 $\a \in \frac{Q}{2}+i\R$ (with $Q=b+b^{-1}$). In order to account for
 the ZZ branes in terms of representation theory, one has to go beyond
 the physical spectrum and take into account the 
degenerate representations appearing
in the Kac table, with momenta
\bea
2\a_{mn}-Q=mb^{-1}+nb \ ,
\eea
where $(m,n)$ are still strictly positive integers, and I ignore the
reflected degenerate representations $2\a_{mn}-Q=-(mb^{-1}+nb)$
because the reflection symmetry of Liouville theory makes them
redundant. Now, the relation (\ref{aj}) between the Liouville momentum
and the $\H$ spin relates the Virasoro degenerate representations to
$\asl$ degenerate representations with spins
\bea
2j_{mn}+1=mb^{-2}+n \ .
\eea
The discrete $AdS_2$ branes should be considered as associated with such
representations, whereas the ordinary $AdS_2$ branes would be
associated with the physical continuous representations $j\in -\frac12+i\R$. This
interpretation of the $AdS_2$ branes was already considered in
\cite{pst01} (section 4.2), which suggested the following relation between
 representation spins $j$ and brane parameters $r$:
\bea
j(r)=-\frac12-\frac{1}{4b^2}+i\frac{r}{2\pi b^2} \ .
\label{jr}
\eea
However, as observed in \cite{pst01}, this relation does not give physical values $j\in
-\frac12+i\R$ for $r$ real due to the term $-\frac{1}{4b^2}$. But this
term precisely corresponds to the shift in the
$AdS_2$-FZZT relation (\ref{spm}), and  now seems rather
natural. The reflection symmetry of the spectrum $j\rar -j-1$ then
corresponds to the invariance of the FZZT branes under $s\rar -s$. 
Now replacing $r$ in eq. (\ref{jr}) with the values
appropriate for discrete $AdS_2$ branes (\ref{aad}) gives the spins of
the $\asl$ degenerate representations with null vector at
nonzero level: $2j(r=i\pi[m-\frac12+nb^2])+1=-(mb^{-2}+n)$. 

There is another series of degenerate representations
of $\asl$ with $m=0$, which do not correspond to Virasoro degenerate
representations because they have a null vector at level zero
\cite{ky92}. D-branes corresponding to these representations are
therefore not expected to be simply related to Liouville theory objects.
There exist natural candidates for such D-branes: 
the $S^2$ branes with imaginary radius 
of \cite{pst01}. In contrast to the $AdS_2$ branes
which preserve an $\SLR$ symmetry out of the $\SLC$ of the $\H$ model,
the $S^2$ branes preserve an $SU(2)$ symmetry. The
degenerate representations with level zero null vectors are unitary as
$SU(2)$ representations, and they indeed appear
in the physical spectrum of the $S^2$ branes. 

The
representations mentioned so far are summarized in the following
table, which should be compared to the picture of the moduli spaces of D-branes in
$\H$ and Liouville theory in the Introduction:
\begin{center}
\psset{unit=.5cm}
\pspicture[](0,-1.3)(10,6.3)
\rput[t](0,6){Virasoro}
\rput[t](10,6){$\asl$}
\psline(-5,5)(15,5)
\rput(0,4){\rnode{FZZT}{\psframebox{$\a\in \frac{Q}{2}+i\R $}}}
\rput(0,2){\rnode{ZZ}{\psframebox{$ 2\a_{mn}-Q=mb^{-1}+nb $}}}
\rput(10,4){\rnode{AdS2}{\psframebox{$j\in -\frac12+i\R$}}}
\rput(10,2){\rnode{AdS2d}{\psframebox{$2j_{mn}+1=mb^{-2}+n  $}}}
 \rput(10,0){\rnode{S2}{\psframebox{$ 2j_n+1=n $}}}
\endpspicture
\end{center}

\subsubsection{Computation of the annulus amplitudes}

In the context of rational conformal field theories, Cardy has shown
that the consistency of the spectrum of open strings on a D-brane
(i.e. the requirement that it consists of finitely many
representations with positive integer multiplicities)
leads to a
strong constraint on the one-point function of that D-brane \cite{car89}. 
In non rational conformal field theories, 
the spectrum of open strings
should consist
of continous states with a positive density and/or discrete states
with positive integer multiplicities. The consistency of the $AdS_2$
branes has already been checked in this way in
\cite{pst01,rib02b}. The study of this type of consistency conditions
is sometimes called the {\it modular bootstrap} approach \cite{zz90}.

The open-string spectrum is related to the one-point function via the
  annulus amplitude  
  $Z^{AdS_2^d}_{(m_1,n_1)(m_2,n_2)}= {\rm
  Tr} \tq^{L_0-\frac{c}{24}}$ where the powers of $\tq$ are the energies of the
  open-string states \footnote{With standard conventions:
  $\tq=\exp-\frac{2\pi i}{\tau}$ and $q=\exp 2\pi i \tau$ where $\tau$
  is the modular parameter of the annulus}.
Like the annulus amplitude for $AdS_2$ branes, the annulus amplitude 
for open strings stretched between two
$AdS_2^d$ branes
is most easily computed in the $\mu$ basis. (A naive $x$-basis
  computation would give a wrong result due to an improper treatment
  of the divergences \cite{pst01}.) 
\bea
Z^{AdS_2^d}_{(m_1,n_1)(m_2,n_2)} &=& \int_{-\frac12+i\R}dj\ \int_\C
\frac{d^2\mu}{|\mu|^2}\ 
\Psi^{AdS_2^d}_{(m_1,n_1)}
\left(\Psi^{AdS_2^d}_{(m_2,n_2)} \right)^*\
\frac{q^{-\frac{b^2}{4}(2j+1)^2}}{\prod_{\ell=1}^\infty (1-q^\ell)^3}
\\
&=& \delta(0)\int_0^\infty 1 \times  \sum_{n\in n_1\times
  n_2} \left(\sum_{m\in m_1\times m_2}+\sum_{m\in (m_1-1)\times
  (m_2-1)} \right) \chi_{mn}(\tq)\ .
\eea
In this formula, $m\in m_1\times m_2$ means $|m_1-m_2|<m<m_1+m_2$ 
  while $m_1+m_2-m$ is an odd integer (like in $\frac{\sin
  m_1x\sin m_2 x}{\sin x}=\sum_{m\in m_1\times m_2} \sin mx$), and
  $\chi_{mn}(q)=\eta^{-3}(q)(q^{-\frac14(mb^{-1}+nb)^2} -q^{-\frac14(mb^{-1}-nb)^2})$
  is an $\asl$ degenerate character \cite{ky92}. The 
  infinite prefactors (which come from the integral $\int_\C d^2\mu$) 
result from the $\SLR$ symmetry of the $AdS_2^d$
  branes and are similar to infinite prefactors appearing in the
  annulus amplitude of $AdS_2$ branes \cite{pst01}. In the case of
  $AdS_2$ branes, there was an extra divergence of the integral $\int dj$ at
  $j=-\frac12$. This zero radial momentum divergence reflected the
  infinite extension of the $AdS_2$ branes in the radial direction and
  is absent in the case of the $AdS_2^d$ branes. 

Therefore, the spectrum of open strings on the $AdS_2^d$ branes is
consistent. The spectrum of open strings between $AdS_2$ and
$AdS_2^d$ branes is also made of discrete states with integer
multiplicities, but these states can have 
imaginary conformal dimensions, as is clear from the formula:
\begin{multline}
Z^{AdS_2-AdS_2^d}_{r,(m,n)} \propto \int dj\
\frac{q^{-\frac{b^2}{4}(2j+1)^2}}{\prod_{\ell=1}^\infty (1-q^\ell)^3} 
\frac{\sin \pi n b^2(2j+1)}{\sin \pi b^2(2j+1)} 
\\
\times \left[ \frac{\sin \pi (m-1)(2j+1)}{\sin \pi (2j+1)}
  \cosh(r-i\frac{\pi}{2})(2j+1) + \frac{\sin \pi m(2j+1)}{\sin \pi (2j+1)}
  \cosh(r+i\frac{\pi}{2})(2j+1) \right] \ .
\end{multline}
The Gaussian integral on $j$ will indeed yield powers of $\tq$ which are
not real. 
In such cases I will say that
$Z^{AdS_2-AdS_2^d}_{r\neq 0,(m,n)}$ has an {\it imaginary spectrum
pathology}. Note however that this pathology is not an inconsistency
of the conformal field theory with boundary conditions defined by
$AdS_2^d$ branes. The pathology only 
prevents the $AdS_2^d$ branes to be interpreted as physical string
theory objects
in the presence of $AdS_2$ branes.

Actually, the ZZ branes with $(m,n)\neq (1,1)$ in Liouville also have
this imaginary spectrum pathology, which does not prevent them from
playing an important r\^ole in the theory.
Note also that
the pathology can be absent in the case of some branes constructed
from the $AdS_2^d$ branes as I will argue in the context of the 2d black hole $\SLU$.

\section{More branes in the 2d black hole}

D-branes in the 2d ``cigar'' Euclidean black hole $\SLU$ can be
obtained from D-branes in the Euclidean $AdS_3$ by a descent procedure
\cite{rs03}. On the one hand this will yield more consistency checks
for the new D-branes constructed in the present article, and on the other
hand this will suggest a comparison with matrix
model results.

\subsection{Known branes and new branes in the 2d black hole}

Let me now recall the one-point functions of the $\SLU$ bulk fields
$\Phi^j_{n',w}$ in the presence of boundary conditions defined by the
D-branes descending from $S^2$ and $AdS_2$ branes in $\H$. 

A D0-brane in the cigar descends from an $S^2$ branes in $\H$ 
labelled by a strictly positive integer $n$:
\bea
\Psi^{D0}_n = \delta_{n'0}\nu_b^{j+\frac12}\frac{k^\frac14
  b^{-\frac12}}{2\pi (-1)^{nw+1}}
\frac{\G(\frac{kw}{2}-j)\G(-\frac{kw}{2}-j)}{\G(-2j)}
\G(1+b^2(2j+1))\sin \pi n b^2(2j+1) \ . 
\label{pd0}
\eea
A D1-brane in the cigar descends from an $AdS_2$ brane in $\H$ with a real
parameter $r$ and an angle $\theta_0$:
\bea
\Psi^{D1}_{r}= 
\delta_{w,0}e^{ i n'\theta_0} \nu_b^{j+\frac12} \frac{ k^{-\frac14}b^{-\frac12}}{2} 
\frac{\Gamma(2j+1) \Gamma(1+b^2(2j+1))}{\Gamma(1+j+\frac{n'}{2})
\Gamma(1+j-\frac{n'}{2})}
\left(e^{-r(2j+1)}+(-1)^{n'} e^{r(2j+1)}
\right) \ .
\label{pd1} 
\eea
A D2-brane in the cigar also descends from an $AdS_2$ brane in $\H$, whose
parameter $r$ now has to be taken pure imaginary $r=i\sigma$. 
The real parameter $\sigma$ of the D2-branes is quantized in units of $2\pi b^2$ and
bounded $|\sigma|<\frac{\pi}{2}(1+b^2)$. 
\begin{multline}
\Psi^{D2}_{\sigma}= \delta_{n',0}\nu_b^{j+\frac12}\frac{ k^{\frac14}b^{-\frac12}}{2\pi} 
 \Gamma(2j+1) \Gamma(1+b^2(2j+1)) 
\\
\times 
\left(\frac{\Gamma(-j+\frac{kw}{2})}{\Gamma(j+1+\frac{kw}{2})}
 e^{i\s(2j+1)}+\frac{\Gamma(-j-\frac{kw}{2})}
  {\Gamma(j+1-\frac{kw}{2})}\ e^{-i\s(2j+1)}\right) \ .
\label{pd2}
 \end{multline}

New discrete branes can be obtained in $\SLU$ from the discrete
$AdS_2$ branes in $\H$. Like the original $AdS_2$ branes, the discrete
$AdS_2$ branes give rise to two families of D-branes in the coset.
Their one-point functions can be obtained from D1- and D2-branes'
one-point functions thanks to the formula (\ref{aad}). Let me first
consider the D1$^d$-branes obtained from the D1-branes:
\begin{multline}
\Psi^{{\rm D1}^d}_{(m,n)}=\delta_{w,0} e^{in'(\theta_0+\frac{\pi}{2})}
\nu_b^{j+\frac12}  2k^{-\frac14}b^{-\frac12}\frac{\G(2j+1)\G(1+b^2(2j+1))}{\G(j+1+\frac{n'}{2})
  \G(j+1-\frac{n'}{2})} 
\\ \times
\sin
\pi\left[(2j+1)(m-\frac12)+\frac{n'}{2}\right] \sin \pi n b^2(2j+1) \ .
\label{pd1d}
\end{multline}
The spectrum encoded in the annulus amplitude $Z^{{\rm
    D1}^d}_{(m_1,n_1)(m_2,n_2)} $ contains a finite number of discrete
    representations with positive integer multiplicities and is
    therefore consistent. 
(However, I did not find the marginal field which
    might have been expected from 
    the existence of a modulus $\theta_0$.) 
Note also that the amplitude $Z^{{\rm D1}-{\rm D1}^d}_{r\neq 0,(m,n)}$
    suffers from the same imaginary spectrum pathology as the
    amplitude $Z^{AdS_2-AdS_2^d}_{r\neq 0,(m,n)}$ in $\H$.  

The D2$^d$-branes obtained from the D2-branes are characterized by the one-point
function:
\begin{multline}
\Psi^{{\rm D2}^d}_{(m,n)} = \delta_{n',0}\nu_b^{j+\frac12}i\frac{ k^{\frac14}b^{-\frac12}}{\pi} 
 \Gamma(2j+1) \Gamma(1+b^2(2j+1))  \sin\pi nb^2 (2j+1) 
\\
\times
\left(\frac{\Gamma(-j+\frac{kw}{2})}{\Gamma(j+1+\frac{kw}{2})}
 e^{i\pi(m-\frac12)(2j+1)}-\frac{\Gamma(-j-\frac{kw}{2})}
  {\Gamma(j+1-\frac{kw}{2})}\ e^{-i\pi(m-\frac12)(2j+1)}\right) \ .
 \label{pd2d}
 \end{multline}
 The spectrum encoded in the annulus amplitude $Z^{{\rm
 D2}^d}_{(m_1,n_1)(m_2,n_2)}$ contains a finite number of discrete
    representations with positive integer multiplicities and is
    therefore consistent. 
 \footnote{The detailed
 computation of this spectrum for $m_1\neq 
 m_2$ would require
 a non-trivial generalization of the calculations in \cite{rs03}. Note
 also that the multiplicities are positive in contrast to the D2-brane case \cite{ipt04},  
due to the sign
 difference between the second lines of eqs (\ref{pd2}) and
 (\ref{pd2d}). }

The spectrum $Z^{{\rm D2}-{\rm D2}^d}_{\sigma, (m,n)} $
 is also consistent and free from the imaginary spectrum
 pathology, because the D2-brane parameter $\sigma$ comes from pure
 imaginary values of the $AdS_2$ brane parameter $r$. However, it
 might be more relevant to examine the amplitude $Z^{{\rm D1}-{\rm
 D2}^d}_{r,(m,n)}$, which is more difficult to compute because of the difference
 in gluing conditions between D1- and D2$^d$-branes. This difficulty
 is no obstacle to finding that $Z^{{\rm D1}-{\rm
 D2}^d}_{\neq 0,(m,n)}$ has the imaginary spectrum
 pathology \footnote{The imaginary
 spectrum pathology is absent from such a discrete annulus amplitude if
 and only if $\Psi^{\rm D1}_r \left(\Psi ^{{\rm D2}^d}_{(m,n)}\right)^*$ is a
 linear combination of a finite number of terms of the type $\cos
 \lambda (2j+1)$ with $\lambda$ either real or pure imaginary. The
 pathology results from such terms with a generic complex $\lambda$.}
except if $(m,n)=(1,1)$, like the amplitude
 $Z^{FZZT-ZZ}_{s,(m,n)}$ in Liouville theory. 
 Notice that $Z^{{\rm D1}-{\rm D0}}_{r\neq 0,n}$ is
 also free from the imaginary spectrum pathology only for $n=1$. The D0-
 and D2$^d$-branes with parameters $n$ and $(1,n)$ respectively behave
 identically in this respect because their overlaps with D1-branes
 only involve closed strings with winding zero, which make no
 difference between them: $\Psi^{\rm DO}_n (w=0)=\Psi^{{\rm
 D2}^d}_{(1,n)}(w=0)$. 

\subsection{Geometric and non-geometric D-branes}

Let me discuss whether the new D1$^d$- and D2$^d$-branes have a
geometric interpretation. A geometric description of the 2d black hole
is possible in the limit $k\rar \infty$ which corresponds to small
string length $\ell_s=\sqrt{\a'}$ (while $\sqrt{k\a '}$ is a fixed
length). 
First recall the geometric interpretation of
the known D0-, D1- and D2-branes \cite{rs03} as zero-, one- and
two-dimensional geometric objects in the 2d black hole. This can be
seen in the large $k$ behaviour of the one-point functions,
\bea
\Psi^{\rm D0}_n \sim k^{-\frac12}  \sp \Psi^{\rm D1}_{(r,\theta_0)} \sim 1 
\sp \Psi^{\rm D2}_{\sigma}\sim k^{\frac12}   . 
\eea
How this behaviour depends on the dimensionality of the D-branes is
indeed consistent with the dependence of the D-branes' tensions
$T\propto (\a')^{-\frac{p}{2}}$ with respect to the D-branes'
dimensions $p$. 

Now the observation (from the previous subsection) that 
closed strings with zero winding couple identically to D0-branes and to D2$^d$-branes
implies that the D2$^d$-branes should be interpreted
as pointlike branes at the tip of the cigar like the D0-branes. The behaviour of
D1$^d$-branes is different:
\bea
\Psi^{{\rm D1}^d}_{(m,n)}\sim k^{-1} \ ,
\eea
thus their one-point functions decrease too fast at large $k$ to allow
a geometric interpretation. It is possible to call the D1$^d$-branes
``anisotropic localized branes at the tip of the cigar'' only in a heuristic
sense. 

\vsp

Let me nevertheless compare this heuristic geometric picture to the
situation in Liouville theory. The localization of the D1$^d$-branes
at the tip of the cigar, and the existence of continuous D1-branes
extending from infinity up to some finite distance from the tip
(a distance determined by their parameter $r$), are similar to the localization
of the ZZ branes at strong Liouville coupling, together with the
existence of FZZT branes extending to infinity. The situation of the
D2$^d$- and D2-branes is quite different since the D2-branes
extend to the tip where the D2$^d$-branes are located. However, a
species of branes with the same gluing conditions as the D2-branes and
a behaviour similar to that of the FZZT branes has been predicted to
exist \cite{fot03,rib03b}: the D-branes descending from $dS_2$ branes
in $AdS_3$ \cite{bp00}, which I will also call $dS_2$ branes: 
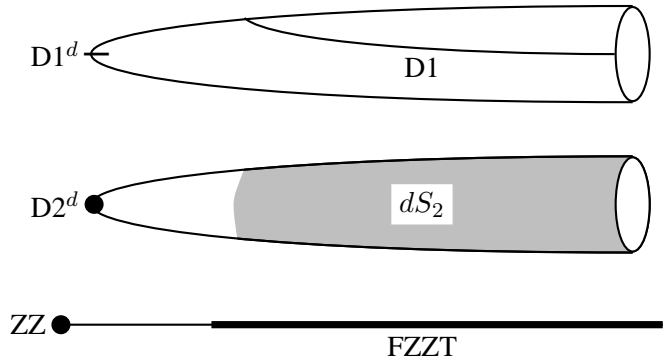
\begin{center}
\psset{unit=.8cm}
\pspicture[](-5,-3)(6,3)
\psellipse(5,2)(.3,.8)
\psbezier(5,2.8)(-7,2.8)(-7,1.2)(5,1.2)
\psbezier(4.7,2)(2,2)(-1,2.2)(-1.45,2.6)
\psdots[dotangle=90,dotstyle=|,dotscale=2](-3.95,2)
\psline(-4.5,-2.5)(5.5,-2.5)
\psdots[dotscale=2](-4.5,-2.5)
\psline[linewidth=.12](-2,-2.5)(5.5,-2.5)
\psellipse(5,-.5)(.3,.8)
\psbezier[fillstyle=solid,fillcolor=lightgray](5,.3)(-7,.3)(-7,-1.3)(5,-1.3)
\psellipse[fillstyle=solid](5,-.5)(.3,.8)
\pscurve[linecolor=white,fillstyle=solid](-1.5,-1.5)(-1.65,-.5)(-1.5,.5)(-4,.5)(-4,-1.5)
\psbezier(5,.3)(-7,.3)(-7,-1.3)(5,-1.3)
\psdots[dotscale=2](-3.95,-.5)
\rput[r](-4.2,2){D1$^d$}
\rput[t](1.5,1.95){D1}
\rput[r](-4.2,-.5){D2$^d$}
\rput*[c](1.5,-.5){$dS_2$}
\rput[r](-4.75,-2.5){ZZ}
\rput[t](1.5,-2.7){FZZT}
\endpspicture
\end{center}
The geometry of the $dS_2$ branes in $AdS_3$ suggests that the $dS_2$
branes in the cigar are parametrized by a real number $r'$ related to
the D2-brane's parameter $\sigma$ by $\sigma=\frac{\pi}{2}+ir'$. The
identifications $\sigma=ir$ and $r= 2\pi b s-i\frac{\pi}{2}$ (from
eq. (\ref{spm})) then imply the relation $r'=2\pi bs$ between the
$dS_2$ brane parameter $r'$ and the FZZT brane parameter $s$. In
addition, the $dS_2$ brane is expected to be invariant under $r'\rar
-r'$ like the FZZT brane under $s\rar -s$. This supports the
idea of a close relationship between $dS_2$ branes in the cigar and FZZT
branes in Liouville theory.

\subsection{A shift equation from $N=2$ Liouville theory}

Let me now compare the D-branes in the 2d black hole with D-branes in
the $N=2$ supersymmetric Liouville theory. The
$N=2$ Liouville theory is indeed equivalent to the $N=2$ supersymmetric 2d
black hole theory \cite{hk01}, which is itself very similar to the bosonic 2d
black hole theory which has been considered in this section. (On the
other hand, the $N=2$ Liouville theory is considerably more complicated than
bosonic Liouville theory.) The comparison of D-branes is relevant to
this article because it will provide an
independent shift equation for the one-point functions of the new
D1$^d$-branes. This is based on the article on $N=2$ Liouville theory
by Hosomichi \cite{hos04}, which among many interesting results
formulates a shift equation with $j$-shift by $\frac{k}{2}$ in addition to
the shift equation with $j$-shift by $\frac12$ considered in subsection
\ref{vse}. These two possible shifts are independent if $k$ is not
rational. However, in contrast to the two elementary $\a$-shifts in Liouville
theory (by $\frac{1}{2b}$ and $\frac{b}{2}$) which are related by
a simple selfduality of the theory, the two shifts in N=2 Liouville
theory must be analyzed independently.

The D1-branes in the 2d black hole (\ref{pd1}) correspond to Hosomichi's 
{\it B-branes} \cite{hos04} (4.55). According to the principles of
subsection \ref{vse}, the D1$^d$-branes should therefore satisfy
\bea
\frac{\Psi^{{\rm D1}^d}_{(m,n)}(j=-\frac{k}{2},n')}{\Psi^{{\rm
      D1}^d}_{(m,n)}(j=0,n'=0) } \overset{!}{=} \left. c^\updownarrow
t^\updownarrow(\frac{n'}{2},-\frac{n'}{2})\right|_{r=i\pi(m-\frac12\pm
  nb^2)}\ ,
\eea
where $c^\updownarrow t^\updownarrow$ is explicitly known \cite{hos04}
(4.55), and the $N=2$ Liouville degenerate spin $\frac{k}{2}$ becomes $-\frac{k}{2}$
 in $\SLU$ after $k\rar k-2$ and reflection.
If proper care is taken of the other differences of conventions, this
equation is found to hold. 

The D2-branes in the 2d black hole (\ref{pd2}) correspond to
Hosomichi's {\it chiral or anti-chiral A-branes} \cite{hos04} (3.26). The
$\frac{k}{2}$-shift equation for these branes \cite{hos04} (4.33) has
a vanishing left-hand side, leading to the condition:
\bea
\frac{\Psi^{{\rm D2}^d}_{(m,n)}(j=-\frac{k}{2},w)}{\Psi^{{\rm
      D2}^d}_{(m,n)} (j=0,w=0)} \overset{!}{=} 0.
\eea 
Surprisingly, this equation holds due to the denominator being
infinite. It therefore provides a rather trivial check of 
the discrete D2-branes's one-point function $\Psi^{{\rm
    D2}^d}_{(m,n)}$. 

To summarize, translating the new $AdS_2^d$ branes to the
D1$^d$-branes in the 2d black hole and then to $N=2$ Liouville theory
has yielded a strong independent check of their consistency. 

In
addition, the new $AdS_2^d$ branes translate into two new families of
discrete D-branes in N=2 Liouville theory, associated to the
continuous {\it B-branes} and {\it chiral or anti-chiral A-branes} of
\cite{hos04}. Note in particular that the N=2 Liouville incarnation of
the D2$^d$-branes differ from the already known {\it non-chiral
degenerate A-branes} 
\cite{hos04} (3.23). These discrete A-branes are actually 
 associated to the continuous {\it
  non-chiral non-degenerate A-branes} \cite{hos04} (3.21). Since there
 exist two types of continuous A-branes in N=2 Liouville theory
 (chiral or anti-chiral on the one hand, non-chiral on the other
 hand), it is not surprising that there exist  two corresponding types of
 discrete A-branes. 

For completeness, let me point out that the
{\it degenerate chiral A-branes} \cite{hos04} (3.33) and their special case
the {\it identity A-brane} \cite{hos04} (3.18) clearly correspond to
D0-branes in the 2d black hole. It would be interesting to 
study the completeness of D-branes in
the 2d black hole and in N=2 Liouville theory.



\acknowledgments{I am supported by a fellowship from the Alexander von Humboldt Stiftung.
I am grateful to Sergei Alexandrov, Thomas Quella, Andreas Recknagel,
Volker Schomerus and J\"org Teschner 
for interesting conversations. In addition, I wish to thank Sakura
Sch\"afer-Nameki and Volker Schomerus for helpful comments on the
draft of this article.
I benefitted from the hospitality of
the Erwin-Schr\"odinger Institut in Vienna. 
}







\providecommand{\href}[2]{#2}\begingroup\raggedright\endgroup

\end{document}